\documentclass[12pt, prd, showpacs]{revtex4}
\usepackage{amsmath}

\input{tcilatex}

\begin{document}

\title{Black hole masking and black hole thermodynamics}
\author{Oleg B. Zaslavskii}
\affiliation{Department of Physics and Technology, Kharkov V.N. Karazin National
University, 4 Svoboda Square, Kharkov, 61077, Ukraine}
\email{zaslav@ukr.net}

\begin{abstract}
Masking of black holes means that, for given total mass and Hawking
temperatures, these data may correspond to either "pure" black hole or a
black hole of a lesser mass surrounded by a massive shell. It is shown that
there is one-to one correspondence between this phenomenon and
thermodynamics of a black hole in a finite size cavity: masking of black
holes is possible if and only if there exists at least one locally unstable
black hole solution in the corresponding canonical ensemble.
\end{abstract}

\keywords{black holes, canonical ensemble, masking}
\pacs{04.70Bw, 04.20.Gz, 04.40 Nr}
\maketitle



\section{Introduction}

In recent years, the phenomenon of mimicking black holes attracts the
particular attention. It implies that, even with reliable observational data
at hand, two quite different types of objects are compatible with a given
set of data. An observer should make a choice between a true black hole or a
compact body with a size slightly bigger than the gravitational radius \cite%
{abr} (see also \cite{mim} and references cited there). For a remote
observer, they reveal themselves almost indistinguishable gravitationally,
although in the vicinity of a (quasi)horizon the difference becomes crucial.
Meanwhile, there is also another phenomenon when one is led to choose not
between a black hole and its mimicker but between different types of black
hole configurations. Namely, as was shown in \cite{ber}, the measurement of
the total mass and (supposing that an observer can measure such things in
principle) Hawking temperature leaves an uncertainty. It is impossible to
learn whether one deals with a "pure" black hole or a black hole of \ a
lesser mass plus a massive shell. Such a phenomenon was called in \cite{ber}
"masking". It is just phenomenon which will be discussed below.

The aim of the present Letter is to draw attention that, actually, there is
close correspondence between two seemingly different areas - masking black
holes from the viewpoint of \ a distant observer and thermodynamics of black
holes enclosed in a finite size cavity. Although our system corresponds to
the microcanonical ensemble (the total mass is fixed, while the Hawking
temperature for a given mass can be calculated), the phenomenon of masking
has roots in the properties of the canonical ensemble. Thus, (i) two quite
different phenomenon (black hole masking and finite size thermodynamics of
black holes) turn out to be mutually connected, (ii) the complementarity
between two quite different types of the gravitational ensemble reveals
itself.

\section{Basic formulas}

Consider a shell around a spherically symmetric black hole, so that the
metric reads 
\begin{equation}
ds^{2}=-A^{2}U_{1}dt^{2}+\frac{dr^{2}}{V_{1}}+r^{2}d\omega ^{2}\text{, }%
r_{+}\leq r\leq R,  \label{1}
\end{equation}%
\begin{equation}
ds^{2}=-U_{2}dt^{2}+\frac{dr^{2}}{V_{2}}+r^{2}d\omega ^{2}\text{, }r\geq R%
\text{.}  \label{2}
\end{equation}%
We do not specify the explicit form of the metric. We only assume that $%
r_{+} $ is the horizon, so $U_{1}(r_{+})=0=V_{1}(r_{+})$. In particular, it
can be the Schwarzschild, Reissner-N\"{o}rdstrom or Scharazchild - de Sitter
metric, etc. We also suppose that inside and outside the type of matter or
fields is the same, so the metric functions differ by the constants of
integration only:%
\begin{equation}
U_{1}=U(r,m_{1})\text{, }U_{2}=U(r,m_{2})\text{, }V_{1}=V(r,m_{1})\text{, }%
V_{2}=V(r,m_{2})\text{.}  \label{uv}
\end{equation}%
Here, $m_{1}=\frac{r_{+}}{2}$, $m_{2}$ is the total ADM mass of the system, 
\begin{equation}
m_{2}=m_{1}+m_{s}\text{,}  \label{m}
\end{equation}%
$m_{s}$ is the contribution to the ADM mass from the shell. As we consider
static systems, we assume that $R>2m_{2}$. The metric can depend on other
independent parameters which are supposed to be the same inside and outside.
For brevity, we omit such a dependence in formulas.

In the state of thermal equilibrium, the system is characterized by the
constant temperature parameter $T_{0}$ having the meaning of the temperature
measured by an observer at infinity. On the shell, the local Tolman
temperature%
\begin{equation}
T=\frac{T_{0}}{\sqrt{U_{2}(R)}}.  \label{tol2}
\end{equation}

The continuity of the metric induced on $r=R$ requires that 
\begin{equation}
A=\sqrt{\frac{U_{2}(R)}{U_{1}(R)}}.  \label{a}
\end{equation}

Then, the temperature $T_{0}$ is equal to \cite{page}, \cite{ber}%
\begin{equation}
T_{0}=AT_{H}^{(1)}  \label{ta}
\end{equation}%
where 
\begin{equation}
T_{H}^{(1)}=\frac{\sqrt{U_{1}^{\prime }(r_{+})V_{1}^{\prime }(r_{+})}}{4\pi }%
=T_{H}(m_{1})
\end{equation}%
is the Hawking temperature calculated for the black hole metric which is
obtained from (\ref{1}) by omitting the screening factor $A$. It follows
from (\ref{a}) - (\ref{ta}) that%
\begin{equation}
T=\frac{T_{H}^{(1)}}{\sqrt{U_{1}(R)}}\text{.}  \label{tol1}
\end{equation}

Eqs. (\ref{tol2}), (\ref{tol1}) express the condition of the thermal
equilibrium: local temperatures calculated from both sides of the shell
coincide.

\section{Masking and interplay between microcanonical and canonical ensembles%
}

Up to now, we simply used general formulas which follow from thermodynamics
and the conditions of matching two metrics. The phenomenon of masking arises
if, additionally, the temperature at infinity can be interpreted as the
Hawking temperature of a black hole without a shell but having the same
total mass:%
\begin{equation}
T_{0}=T_{H}(m_{2})  \label{tm2}
\end{equation}%
where $m_{2}$ is its ADM mass coinciding with (\ref{m}).

For simplicity, we assume that the shell does not carry an electric charge
and is characterized by the radius and mass only. Then, it follows from (\ref%
{tol2}) and (\ref{tm2}) that 
\begin{equation}
T=f(m_{1},R)=f(m_{2},R)  \label{td}
\end{equation}%
where $f(m,R)=\frac{T_{H}(m)}{\sqrt{U(R.m)}}$.

In general, we can suppose that eq. (\ref{td}) has a set of roots $m_{i}$
where $i=1,2,...N$.

Now, the key observation consists in that eq. (\ref{td}) arises in the
canonical ensemble. Namely, if we consider a cavity of the areal radius $R$
and fix the local temperature $T$ on its boundary, the equation%
\begin{equation}
T=f(R,m)  \label{tr}
\end{equation}%
determines masses of a black hole which can exist inside \cite{y}. For the
Schwarzschild case, if the temperature is high enough, there are two roots,
one unstable $m_{1}$ and the other - stable with $m_{2}>m_{1}$ (see \cite{y}
for details and discussion below). In general, there exists a number of
roots $m_{i}$, $i=1,2...N$. In what follows we shall also \ make a
physically reasonable assumption that the energy is a monotonically
increasing function of the ADM mass, $\frac{dE}{dm}>0$. Then, we have the
following

\textit{Theorem}. Masking of black holes is possible if and only if there
exists at least one locally unstable black hole solution in the
corresponding canonical ensemble.

\textit{Proof.} \ (i) Let us suppose that masking is possible and there are
no unstable roots. This means that the heat capacity $C$\thinspace $=\frac{dE%
}{dT}>0$. As $\frac{dE}{dm}>0$, we see that $\frac{dm}{dT}>0$, so $f(m,R)$
is the monotonic function of $m$. Hence, eq. (\ref{tr}) has only one root
and masking is impossible, so we obtained the contradiction. Thus, if there
is masking, unstable roots are inevitable.

(ii) In a similar manner, one can prove that the reverse is also true. Let
locally unstable roots do exist. Then, the function $f(m)$ has at least one
segment with $\frac{\partial f}{\partial m}<0$. From the other hand, for $R$
close to $r_{+}$, the hole occupies almost all the cavity, $U\rightarrow 0$, 
$f\rightarrow \infty $, so the function $f$ increases when $r_{+}$
approaches $R$. In doing so, the contribution to the mass from the matter
between the gravitational radius and the boundary is negligible, so $%
m\approx \frac{r_{+}}{2}\rightarrow \frac{R}{2}$. Thus, there exist the
branch with $\frac{\partial f}{\partial m}>0$. Between both branches there
is at least one local minimum. Taking the value of $T$ bigger than this
minimum (but smaller than the maximum value of $f$ nearest to it from the
left if such a maximum exist), we obtain the solution with at least two
different roots, so that masking is indeed possible. Thus, the allowed range
of solutions falls into the interval determined by the inequality%
\begin{equation}
f(m,R)\geq f_{0}(R)  \label{f0}
\end{equation}%
where $f_{0}(R)=f(m_{0}(R),R)$ is the minimum of $f$, $\left( \frac{\partial
f}{\partial m}\right) _{m=m_{0}(R)}=0$.

It is worth stressing that the correspondence which we established relates
two different types of systems. Type 1 implies that the space-time is
asymptotically flat, with the total mass $m_{2}$ fixed that corresponds to
the \textit{microcanonical} ensemble. Type 2 represents the \textit{%
canonical }ensemble where the physical manifold is restricted by inequality $%
r\leq R$ that represents the interior of the finite size cavity, but there
is no an external remote observer and there is no shell at all. In a sense,
we have a complementarity of two ensembles and of two types of boundaries
(the thing shell in an infinite space and the boundary enclosing the
physical manifold).

\section{Example: Schwarzschild black hole}

Now I illustrate the above consideration using the Schwarzschild black hole
as an example. The possibility of masking such a black hole was pointed out
in \cite{ber}, thermodynamics of finite size system for such black holes was
considered in \cite{y} but now my aim is to compare both phenomena. Now, in
geometric units $T_{H}=(8\pi m)^{-1}$, $U=V=1-\frac{2m}{r}$, \thinspace $%
f=8\pi m\sqrt{1-\frac{2m}{R}\text{ }}$, $E=R(1-\sqrt{V(R)})$. The condition
of masking (\ref{tm2}), (\ref{td}) can be written in the form%
\begin{equation}
\frac{1}{8\pi m_{2}\sqrt{1-\frac{2m_{2}}{R}}}=\frac{1}{8\pi m_{1}}\frac{1}{%
\sqrt{1-\frac{2m_{1}}{R}}}  \label{s}
\end{equation}%
which coincides with eq. (3.2) of \cite{ber}. This exactly corresponds also
to the equation that, for a fixed local temperature $T$ on the boundary,
defines possible values of the black hole mass inside the cavity:%
\begin{equation}
T=f(m,R)=(8\pi m\sqrt{1-\frac{2m}{R}})^{-1}.  \label{rt}
\end{equation}

For a given $R$, the quantity $f(m,R)$ as the function of $m$ (or the
gravitational radius $r_{+}=2m$) has two branches - monotonically decreasing
and monotomically increasing ones which meet in the point of the minimum of $%
f$ where $m=\frac{R}{3}$. For given $T,R$ the solutions of this equation
exist for $RT>\frac{\sqrt{27}}{8\pi }$. There are two roots with masses $%
m_{1}<m_{2}$, the light one on the decreasing branch of $f$ is locally
unstable, the heavy one on the increasing branch of $f$ is locally stable
(see \cite{y} for details).

To render the aforementioned inequality in terms of mass (thus translating
the properties of the canonical ensemble to those of the microcanonical
one), it is necessary to find the minimum of $f(m,R)$ with respect to $m$
for a given $R$. The solution lies above such a minimum. Then, it is easy to
show that $R<3m_{2}$. In combination with the condition $R>2m_{2}$, we
obtain the restriction%
\begin{equation}
2m_{2}<R<3m_{2}\text{.}  \label{3r}
\end{equation}

This equation is a particular case of eq. (\ref{f0}), now $f_{0}=\frac{3%
\sqrt{3}}{8\pi R}$. Thus, any "pure" black hole with a given mass $m_{2}$
has an infinite set of "doubles" obtained with the help of the shell of
different radii in the interval (\ref{3r}). Vice versa, if from the very
beginning we take a black hole surrounded by a shell of the areal radius $R$%
, such a configuration has a pure black hole as its double. The total mass
should lie in the interval $\frac{R}{3}<m_{2}<\frac{R}{2}$. If the masses of
the black hole and shell are also fixed, not any configuration can be masked
since for $m_{2}>\frac{R}{3}$ inequality (\ref{3r}) is violated. If it is
satisfied, an observer at infinity cannot distinguish between two
configurations - the black hole without the shell having the mass $m_{2}$
and the black hole with the mass $m_{1}$ surrounded by the shell of the mass 
$m_{s}=m_{2}-m_{1}$. Both configurations have the same total ADM mass $m_{2}$
and the same Hawking temperature measured at infinity $T_{H}=\frac{1}{8\pi
m_{2}}$.

If one places the shell at $R=3m_{2}(1-\delta )$, $\delta \ll 1$, it follows
from (\ref{s}) that $m_{1}\approx \frac{R}{3}(1-\delta )$, $m_{s}\approx 
\frac{2R}{3}\delta $, so in the limit $\delta \rightarrow 0$ the effect of
masking almost vanishes. If one considers the shell at $R=2m_{2}(1+%
\varepsilon )$, $\varepsilon \ll 1$, it turns out that $m_{1}\approx \frac{1%
}{8\pi T}\approx m_{2}\sqrt{\varepsilon }\ll m_{2}$, $m_{s}\approx m_{2}$.
Thus, the mass stems almost entirely from the shell, whereas the
contribution of the black hole inside the shell becomes negligible. In this
respect, such a situation is close to that for black hole mimickers where
there is no a black hole at all, the size of the body approaching the
gravitational radius. In doing so, large tangential stresses develop on the
shell to maintain it in equilibrium \cite{mim} but they are irrelevant for a
distant observer.

\section{Summary}

In general, the canonical ensemble implies that the system is enclosed
inside some cavity of the finite size, so the region with $r>R$ is not part
of the physical system at all. By contrary, the phenomenon of masking
implies that measurement are done at infinity. Nonetheless, it turned out
that these so different (in a sense, mutually complimentary) phenomena are
intimately tied. The effects of finite size in black hole thermodynamics 
\cite{y} are sometimes considered as a pure academic matter having no
observational consequences. However, if we assume that the Hawking radiation
is detectable, properties of finite size black hole thermodynamics should be
taken into account just in observations to single out the potential effect
of black hole masking.


\begin{thebibliography}{9}
\bibitem{abr} M. A. Abramowicz, W. Kluzniak, and J. P. Lasota, Astron.
Astrophys. 396 (2002) L31.

\bibitem{mim} J. P. S. Lemos and O. B. Zaslavskii, Phys. Rev. D 78 (2008)
024040.

\bibitem{ber} V. A. Berezin and A. Smirnov, Grav. Cosmol. 13 (2007) 43.

\bibitem{page} P. C. W. Davies, L. H. Ford, and D. N. Page, Phys. Rev. D 34
(1986) 1700.

\bibitem{y} J. W. York, Jr. Phys. Rev. D 33 (1986) 2092.
\end{thebibliography}
\end{document}